# Separability and entanglement of two qubits density matrices using Lorentz transformations

Y. Ben-Aryeh[*] and A. Mann[†]

*Physics Department, Technion-Israel Institute of Technology,
Haifa 32000, Israel*

[*] phr65yb @ physics.technion.ac.il   ;   [†] ady @ physics.technion.ac.il

**Abstract**
Explicit separability of general two qubits density matrices is related to Lorentz transformations. We use the 4-dimensional form $R_{\mu,\nu}$ ($\mu, \nu = 0,1,2,3$) of the Hilbert-Schmidt (HS) decomposition of the density matrix. For the generic case in which Lorentz transformations diagonalize $R_{\mu,\nu}$ (into $s_0$, $s_1$, $s_2$, $s_3$) we give relations between the $s_\mu$ and the $R_{\mu,\nu}$. In particular we consider two cases: a) Two qubits density matrices with one pair of linear terms in the HS decomposition. b) Two qubits density matrices with two or three symmetric pairs of linear terms. Some of the theoretical results are demonstrated by numerical calculations. The four non-generic cases (which may be reduced to case $a$) are analyzed and the non-generic property is related explicitly to Lorentz velocity $\beta = 1$ which is not reachable physically.

Condensed paper title: Lorentz transformations of 2 qubits.

## 1. Introduction

In the case of two qubits states there is a simple condition for separability and entanglement. According to the Peres-Horodecki (P-H) criterion[1,2] if the partial transpose (PT) of two qubits state leads to negative eigenvalues of the PT matrix $\rho_{AB}(PT)$, then the density matrix is entangled, otherwise it is separable. The general two qubits density matrix in the Hilbert-Schmidt (HS) decomposition is:

$$4\rho = I_A \otimes I_B + \vec{a} \cdot \vec{\sigma}_A \otimes I_B + I_A \otimes \vec{b} \cdot \vec{\sigma}_B + \sum_{l,m=1}^{3} t_{l,m} \sigma_{l,A} \otimes \sigma_{m,B} \quad . \tag{1.1}$$

$\vec{\sigma}_{l,A}$ and $\vec{\sigma}_{m,B}$ represent Pauli spin matrices of the qubits $A$ and $B$, respectively. $\vec{a}$ and $\vec{b}$ are 3-dimensional vectors and $I_A$, and $I_B$ are $2 \times 2$ unit matrices. We denote $\vec{a} \cdot \vec{\sigma}_A \otimes I_B$ and $I_A \otimes \vec{b} \cdot \vec{\sigma}_B$ as the linear terms, A and B respectively. The number of parameters describing the two-qubits density matrices can be reduced by local transformations.[3–12] We consider $\rho$ and $\rho_M$ to be of the same equivalence class when

$$\rho \to \rho^M = M \rho M^\dagger \quad , \quad M = M_A \otimes M_B \quad , \tag{1.2}$$

with $M_A$ and $M_B$ invertible. Such equivalence preserves the positivity and separability of the density matrices. Using the singular value decomposition (SVD) the density matrix can be transformed to

$$4\rho = I_A \otimes I_B + \vec{a} \cdot \vec{\sigma}_A \otimes I_B + I_A \otimes \vec{b} \cdot \vec{\sigma}_B + \sum_{i=1}^{3} t_i \sigma_{i,A} \otimes \sigma_{i,B} \quad , \tag{1.3}$$

where $t_i$ are the singular values of the matrix $t_{l,m}$. We denote the last term by T.

In our previous work[10] we studied the explicit constructions of separable two qubits density matrices based on the study of Lorentz transformations developed by Verstraete et al.[4–6] Following this approach an arbitrary two qubits state can be written in the form

$$\rho = \frac{1}{4} \sum_{\mu,\nu=0}^{3} R_{\mu,\nu} (\sigma_\mu)_A \otimes (\sigma_\nu)_B \quad , \tag{1.4}$$

where $\sigma_0 = I$, $\sigma_i$, $i = 1,2,3$ are Pauli matrices. A matrix $\rho$ corresponding to a quantum state is Hermitian, positive definite and has a unit trace, where unit trace of $\rho$ can be obtained by proper normalization. The Hermitcity of $\rho$ is equivalent to the condition that the matrix $R = [R_{\mu,\nu}]$ is real. Detailed analysis for the two qubits density matrix which is of the form (1.4) has been given by Caban et al.[11,12] Verstraete et al.,[4–6] have shown that the $4 \times 4$ matrix $R_{\mu,\nu}$ can be written as

$$R = L_1 \Sigma L_2^t \quad . \tag{1.5}$$

Here $L_1$ and $L_2$ are proper local Lorentz transformations, and $\Sigma$ is either the diagonal form $\Sigma = diag(s_0, s_1, s_2, s_3)$ (the generic case[5]) or of four special other specific forms[5] (the non-generic cases for which $R$ cannot be diagonalized by Lorentz transformations). In the present work we give explicit Lorentz transformations diagonalizing $R$ for some generic cases. Also we show explicitly in the non-generic cases how the Lorentz transformation cannot diagonalize $R$.

It has been shown in various works [10,13–15] that for the two qubits density matrix given by (1.3) under the condition that $\vec{a} = \vec{b} = 0$ (referred to as density matrices with maximally disordered subsystems (MDS), [13,10,16,17] a necessary and sufficient condition for separability is given by

$$\sum_{i=1}^{3} |t_i| \leq 1 \quad . \tag{1.6}$$

We prove here first the following Theorem: For the density matrix of Eq. (1.3) the condition of Eq. (1.6) is necessary for separability, but it is not sufficient. Proof: If $\rho$ is separable then $\rho^t$ (the transpose of $\rho$) is also separable. A unitary transformation of $\rho^t$ around $y$ by $180^0$ yields $\rho^{tu}$, which is also separable. Explicitly, we obtain

$$4\rho^{tu} = I_A \otimes I_B - \vec{a} \cdot \vec{\sigma}_A \otimes I_B - I_A \otimes \vec{b} \cdot \vec{\sigma}_B + \sum_{i=1}^{3} t_i \sigma_{i,A} \otimes \sigma_{i,B} \quad . \tag{1.7}$$

From equations (1.3) and (1.7) we get:

$$4\tilde{\rho} \equiv \frac{1}{2}(4\rho + 4\rho^{tu}) = I_A \otimes I_B + \sum_{i=1}^{3} t_i \sigma_{i,A} \otimes \sigma_{i,B} \quad . \tag{1.8}$$

Since $\rho$ and $\rho^{tu}$ are separable $\tilde{\rho}$ is separable too. Hence we get from Eq. (1.6) $\sum_{i=1}^{3} |t_i| \leq 1$.

Therefore the condition $\sum_{i=1}^{3} |t_i| \leq 1$ in Eq. (1.3) is a necessary condition for separability, but it is not sufficient. We demonstrate the insufficiency by using the following example: Let us assume $t_1 = t_2 = t_3 = 0.3$; $a_2 = b_2 = 0.64$; $a_1 = b_1 = a_3 = b_3 = 0$, then the corresponding $\rho$ is a density matrix as all its eigenvalues are positive, given approximately by:

$$4\lambda_1 = 2.58, \quad 4\lambda_2 = 0.02, \quad 4\lambda_3 = 0.1, \quad 4\lambda_4 = 1.30 \quad . \tag{1.9}$$

But by carrying the PT transformation relative to qubit A (PTUA) we get a negative eigenvalue as: $4\lambda_1(PTA) = 2.715, \quad 4\lambda_2(PTA) = -0.115, \quad 4\lambda_3(PTA) = 0.70, \quad 4\lambda_4(PTA) = 0.70$. (1.10)

So, we demonstrated here that this $\rho$ is not separable although $\sum_{i=1}^{3} |t_i| \leq 1$.

We can separate the density matrix (1.3) into 4 parts as follows:

$$4\rho = I_A \otimes I_B + A + B + T \; ; \; A = \vec{a} \cdot \vec{\sigma}_A \otimes I_B \; ; \; B = I_A \otimes \vec{b} \cdot \vec{\sigma}_B \; ; \; T = \sum_{i=1}^{3} t_i \sigma_{i,A} \otimes \sigma_{i,A} \quad . \tag{1.11}$$

The PT of the density matrix $\rho$ is equivalent to the change $\sigma_y \to -\sigma_y$ of one of the two qubits (say A), leaving $\sigma_{x,A}, \sigma_{z,A}$ unchanged. A unitary transformation of this qubit around $y$ by $180^0$ does not change $\sigma_y$ but leads to the changes: $\sigma_x \to -\sigma_x$ ; $\sigma_z \to -\sigma_z$. For the special case for which $B=0$ we perform the PTU transformation (PT plus the unitary transformation) relative to A (PTUA):

$$4\rho(PTUA) = I_A \otimes I_B - A - T \quad . \tag{1.12}$$

Then, for this special case we get:

$$\frac{4\rho + 4\rho(PTUA)}{2} = I_A \otimes I_B \to \lambda_i(\rho;PTUA) = \frac{1}{2} - \lambda_i(\rho) \ (i=1,2,3,4) \ . \tag{1.13}$$

Then we get that a necessary and sufficient condition for separability in this case is given by:

$$\lambda_i(\rho) \ (i=1,2,3,4) \leq \frac{1}{2} \quad . \tag{1.14}$$

Eq. (1.14) was noted earlier for MDS density matrices (A=B=0)[13,10] but is valid also for the special cases for which: $B=0, A \neq 0$ or $B \neq 0, A = 0$

By using the 4-dimensional form (1.4) for the two-qubits density matrix (1.3) $R$ is given as

$$R = \begin{pmatrix} 1 & a_1 & a_2 & a_3 \\ b_1 & t_1 & 0 & 0 \\ b_2 & 0 & t_2 & 0 \\ b_3 & 0 & 0 & t_3 \end{pmatrix} \quad . \tag{1.15}$$

In the present analysis we study the use of Lorentz transformations for simplifying the density matrix of Eq. (1.3) by eliminating the linear terms (which include the parameters $a_i$ and $b_i$) and transforming $R$ into diagonal form for the generic case described by Verstraete et al.[5]

For the generic case, diagonal $\Sigma$ [5], an explicitly separable expression for separable two qubits states was given in our previous work[10] (Eq. (27)), as a function of the parameters $s_\mu$. The necessary and sufficient condition for separability is given by

$$|s_1| + |s_2| + |s_3| \leq s_0 \quad . \tag{1.16}$$

While this condition was derived by Verstraete et al.,[5] they have not given the explicit relations between the parameters $a_i, b_i, t_i \ (i=1,2,3)$ and the parameters $s_\mu (\mu=0,1,2,3)$. In the present work we study such explicit relations. We study especially two cases: a) when we have one pair of linear parameters $a_i$ and $b_i$ (i=1 or 2 or 3). b) The case in which we have 2 or 3 symmetric parameters $a_i = b_i (i = 1, 2, 3)$. A similar analysis can be made for more general cases but the analysis turns out to be much more complicated.

We can reformulate the condition given by Verstraete et al,[5] for separability in a normalized form where in the generic case, $R$ of Eq. (1.15) is transformed by Lorentz transformations into a normalized form:

$$R_L = \begin{pmatrix} 1 & 0 & 0 & 0 \\ 0 & t'_1 & 0 & 0 \\ 0 & 0 & t'_2 & 0 \\ 0 & 0 & 0 & t'_3 \end{pmatrix} . \quad (1.17)$$

Here $t'_1 = \frac{s_1}{s_0}$ ; $t'_2 = \frac{s_2}{s_0}$ ; $t'_3 = \frac{s_3}{s_0}$ . Then, the condition for separability in the generic case becomes:

$$\sum_{i=1}^{3} |t'_i| \leq 1 . \quad (1.18)$$

The effects of the linear terms in the present generic case, is to replace the condition (1.6) for the two qubits MDS density matrix by condition (1.18). While Eq. (1.18) seems to be quite simple the explicit calculations of the parameters $t'_i$ $(i = 1, 2, 3)$ in the generic case by Lorentz transformations turns to be very complicated and we give explicit calculations for some important cases. We give also some comments on the non-generic cases in Section 2.

## 2. Lorentz transformations for the density matrix of a two qubits state including only two linear terms: $a_1$ and $b_1$

We make the analysis for the case which we have the linear terms $a_1$ and $b_1$. (A similar analysis can be made if we have the terms $a_2$ and $b_2$, or $a_3$ and $b_3$). This simple case has the advantage that it can be solved analytically and also all the non-generic cases considered by Verstraete et al.,[4–6] and Caban et al.,[11,12] are reducible to special cases of choices of these parameters.

We make the analysis for $R$ given by

$$R = \begin{pmatrix} 1 & a_1 & 0 & 0 \\ b_1 & t_1 & 0 & 0 \\ 0 & 0 & t_2 & 0 \\ 0 & 0 & 0 & t_3 \end{pmatrix} \qquad (2.1)$$

Then, the density matrix corresponding to the matrix $R$ of Eq. (2.1) can be written as

$$\rho = \frac{1}{4}\left[(I)_A \otimes (I)_B + a_1(\sigma_x)_A \otimes (I)_B + b_1(I)_A \otimes (\sigma_x)_B + \sum_{i=1}^{3} t_i(\sigma_i)_A \otimes (\sigma_i)_B\right] \qquad (2.2)$$

We study the effects of one pair of linear terms on the density matrix and its separability. The 4 eigenvalues of the density matrix (2.2) are:

$$4\lambda_{1,2} = 1 + t_1 \mp \sqrt{(a_1+b_1)^2 + (t_2-t_3)^2} \quad ; \quad 4\lambda_{3,4} = 1 - t_1 \mp \sqrt{(a_1-b_1)^2 + (t_2+t_3)^2} \, . \qquad (2.3)$$

One should notice that if we change the parameters $a_1$ and $b_1$ to $a_2$ and $b_2$ or to $a_3$ and $b_3$ we get the same eigenvalues under the transformations $1 \rightarrow 2, 3$, respectively. In order to use the Lorentz transformations we transform $R$ as:

$$R_L = L_A R L_B \quad ; \quad \gamma_A = \frac{1}{\sqrt{1-\beta_A^2}} \quad ; \quad \gamma_B = \frac{1}{\sqrt{1-\beta_B^2}}$$

$$L_A = \begin{pmatrix} \gamma_A & -\gamma_A\beta_A & 0 & 0 \\ -\gamma_A\beta_A & \gamma_A & 0 & 0 \\ 0 & 0 & 1 & 0 \\ 0 & 0 & 0 & 1 \end{pmatrix} \quad ; \quad L_B = \begin{pmatrix} \gamma_B & -\gamma_B\beta_B & 0 & 0 \\ -\gamma_B\beta_B & \gamma_B & 0 & 0 \\ 0 & 0 & 1 & 0 \\ 0 & 0 & 0 & 1 \end{pmatrix} \qquad (2.4)$$

$R_L$ is given by:

$$R_L = \begin{pmatrix} \gamma_A\gamma_B(1 - b_1\beta_A - a_1\beta_B + \beta_A\beta_B t_1) & \gamma_A\gamma_B(-\beta_B + b_1\beta_A\beta_B + a_1 - \beta_A t_1) & 0 & 0 \\ \gamma_A\gamma_B(-\beta_A + a_1\beta_A\beta_B + b_1 - \beta_B t_1) & \gamma_A\gamma_B(-b_1\beta_B + \beta_A\beta_B - a_1\beta_A + t_1) & 0 & 0 \\ 0 & 0 & t_2 & 0 \\ 0 & 0 & 0 & t_3 \end{pmatrix} \qquad (2.5)$$

In the generic case we can eliminate the non-diagonal matrix elements of $R_L$ by the requirements:

$$-\beta_B + b_1\beta_A\beta_B + a_1 - \beta_A t_1 = 0 \quad ; \quad -\beta_A + a_1\beta_A\beta_B + b_1 - \beta_B t_1 = 0 \quad . \qquad (2.6)$$

The equation on the left hand side of (2.6) multiplied by $a_1$ minus the equation on the right hand side of (2.6) multiplied by $b_1$ gives:

$$a_1^2 - b_1^2 + b_1(\beta_A + \beta_B t_1) - a_1(\beta_B + \beta_A t_1) = 0 \quad . \tag{2.7}$$

From Eq. (2.7) we get the relations:

$$\beta_B = \frac{a_1^2 - b_1^2 + (b_1 - a_1 t_1)\beta_A}{a_1 - b_1 t_1} \quad ; \quad \beta_A = \frac{b_1^2 - a_1^2 + (a_1 - b_1 t_1)\beta_B}{b_1 - a_1 t_1} \quad . \tag{2.8}$$

We substitute the value of $\beta_B$ from Eq. (2.8) into Eq. (2.6) and then we get a quadratic equation for $\beta_A$:

$$(b_1 - a_1 t_1)\beta_A^2 + (a_1^2 - b_1^2 + t_1^2 - 1)\beta_A + (b_1 - a_1 t_1) = 0 \quad . \tag{2.9}$$

The solution for $\beta_A$ is given by

$$\beta_A = \frac{1}{2(b_1 - a_1 t_1)}$$
$$\left[ (1 - t_1^2 - a_1^2 + b_1^2) \pm \sqrt{t_1^4 - 2t_1^2(1 + a_1^2 + b_1^2) + 8a_1 b_1 t_1 + 1 + a_1^4 + b_1^4 - 2a_1^2 b_1^2 - 2(a_1^2 + b_1^2)} \right] \quad . \tag{2.10}$$

$\beta_B$ is given by a similar equation, interchanging $a_1$, $b_1$. Although we get two solutions represented by the $\pm$ sign we need to choose the solution with the minus sign so that $\beta_A$ vanishes according to Eq. (2.9) in the limit $a_1, b_1 \to 0$ (when there is no need for Lorentz transformation).

Two non-generic cases (out of the four described by Verstraete et al.[6]) are given in our notation, in normalized form, as  a)  $a_1 = 1, b_1 = t_1 = t_2 = t_3 = 0$.  b)  $b_1 = 1, a_1 = t_1 = t_2 = t_3 = 0$. These two cases correspond to Caban et al.,[11] Eq. (8). We get from Eq. (2.6), for case  a) $\beta_B = 1$, and for case  b) $\beta_A = 1$ so that for these cases the Lorentz transformations are not reachable physically. These states are given by a pure state of A multiplied by the unit matrix of B and vice versa.

We study further the results for two extreme cases:

**The symmetric case:** $a_1 = b_1 \equiv a$

Under such symmetric condition Eq. (2.10) is reduced to the form:

$$\beta = \beta_A = \beta_B = \frac{1}{2a}\left[1 + t_1 - \sqrt{(1 + t_1)^2 - 4a^2}\right] \quad . \tag{2.11}$$

We find here that $\beta$ will be real under the condition:

$$|1+t_1| \geq |2a| \qquad . \tag{2.12}$$

This condition is always satisfied due to the requirement that the eigenvalue $\lambda_1$ given by Eq. (2.3) is non-negative. For the limiting value $1+t_1 = |2a|$ in Eq. (2.12), this equation leads to the value $\beta = 1$ which physically is unreachable and in this case it follows from the non-negativity of the eigenvalues that $t_2 = t_3$. This case is non-generic,[6] and corresponds to Caban et al.,[11] (Eq. (12)). For comparison with Verstraete et al.,[6] we find that they have two symmetric non generic cases written in our notations : c) $a_1 = b_1 = 1/2$, $t_1 = 0$, d) $a_1 = b_1 = t_1 = 1$. For the latter case the non-negativity of the eigenvalues of $\rho$ imposes the additional condition: $t_2 = t_3 = 0$. This last case is always non-separable as easily proved by the PT transformation. The other non-generic cases are separable.

For the case where $a_1 = b_1 = a$ ($|a| < 1$) under conditions (2.6), $R_L$ of (2.5) becomes (except for the limiting case mentioned above) :

$$R_L = \begin{pmatrix} s_0 = \gamma^2(1-2a\beta+\beta^2 t_1) & 0 & 0 & 0 \\ 0 & s_1 = \gamma^2(-2a\beta+\beta^2+t_1) & 0 & 0 \\ 0 & 0 & s_2 = t_2 & 0 \\ 0 & 0 & 0 & s_3 = |t_3| \end{pmatrix} . \tag{2.13}$$

The necessary and sufficient condition for separability is given by

$$\frac{|s_1|}{s_0} + \frac{|s_2|}{s_0} + \frac{|s_3|}{s_0} \equiv |t'_1| + |t'_2| + |t'_3| = \frac{|-2a\beta+\beta^2+t_1|}{(1-2a\beta+\beta^2 t_1)} + \frac{|t_2|+|t_3|}{\gamma^2(1-2a\beta+\beta^2 t_1)} \leq 1 \qquad . \tag{2.14}$$

By using (2.11) and (2.14) for the example in the introduction, where $a = 0.64$; $t_1 = t_2 = t_3 = 0.3$ we get: $\beta = 0.8382$; $\gamma^2 = 3.3615$; $s_0 = 0.4636$; $s_1 = -0.2390$; $t'_2 = t'_3 = 0.6471$; $t'_1 = \frac{|s_1|}{s_0} = 0.5153$

We get $\sum_{i=1}^{3} |t'_i| = 1.8095 > 1$, so that the density matrix in this example is not separable.

It is interesting to note that under the condition $s_1 = \gamma^2\left(-2a\beta + \beta^2 + t_1\right) \geq 0$ in Eq. (2.13) (which holds for $a^2 \leq |t_1|$ and then we have $\beta = a$) we get $s_0 - s_1 = 1 - t_1$, so in this case $1 - |t_1| - |t_2| - |t_3| \geq 0$ is both necessary and sufficient for separabilty. Changes from this criterion may occur only for negative values of $s_1$.

**The extreme non-symmetric case $b_1 = 0$**

Under the condition: $b_1 = 0$ Eq. (2.10) is reduced to

$$\beta_A = \frac{-1}{2a_1 t_1}\left[\left(1 - a_1^2 - t_1^2\right) - \sqrt{\left(1 - a_1^2 - t_1^2\right)^2 - 4a_1^2 t_1^2}\right] \quad . \tag{2.15}$$

We find here that $\beta_A$ will be real only under the condition:

$$\left|1 - t_1^2 - a_1^2\right| \geq \left|2a_1 t_1\right| \quad . \tag{2.16}$$

For cases in which the condition $b_1 = 0$ is satisfied, the vanishing of the off-diagonal matrix elements of Eq. (2.5) leads to the relation

$$\beta_B = a_1 - \beta_A t_1 \quad . \tag{2.17}$$

By using (2.5) and (2.17) (with $b_1 = 0$) we get:

$$s_1 = \gamma_A \gamma_B \left(\beta_A \beta_B - a_1 \beta_A + t_1\right) = \frac{\gamma_B t_1}{\gamma_A} \quad , \tag{2.18}$$

$$s_0 = \gamma_A \gamma_B \left(1 - a_1 \beta_B + \beta_A \beta_B t_1\right) = \gamma_A \gamma_B \left[1 - \left(a_1 - \beta_A t_1\right)^2\right] = \frac{\gamma_A}{\gamma_B} \quad . \tag{2.19}$$

It follows that for $b_1 = 0$ the Lorentz transformation (2.5) transforms $R$ into the form:

$$R_L = \begin{pmatrix} \frac{\gamma_A}{\gamma_B} & 0 & 0 & 0 \\ 0 & \frac{\gamma_B t_1}{\gamma_A} & 0 & 0 \\ 0 & 0 & t_2 & 0 \\ 0 & 0 & 0 & t_3 \end{pmatrix} \quad . \tag{2.20}$$

The necessary and sufficient condition for separability is then given by

$$\frac{|s_1|}{s_0} + \frac{|s_2|}{s_0} + \frac{|s_3|}{s_0} \equiv |t'_1| + |t'_2| + |t'_3| = |t_1|\left(\frac{\gamma_B}{\gamma_A}\right)^2 + \left(|t_2| + |t_3|\right)\frac{\gamma_B}{\gamma_A} \leq 1 \quad . \tag{2.21}$$

Let us examine the necessary and sufficient condition for separability in an example where: $a_1 = 0.2; t_1 = t_2 = t_3 = 0.3; b_1 = 0$. For this case we get:

$$|t'_1| + |t'_2| + |t'_3| = 0.9169 < 1 \quad . \tag{2.22}$$

So, this $\rho$ is separable.

Since the case $b_1 = 0$ corresponds to the analysis made after Eq. (1.12) for $B = 0$, we can use Eq. (1.14) as a sufficient and necessary condition for separability. The eigenvalues in this example are given by $4\lambda_{1,2} = 1 + 0.3 \mp 0.2$ ; $4\lambda_{3,4} = 1 - 0.3 \mp \sqrt{0.4}$. As all eigenvalues in this example are positive and less than $\frac{1}{2}$ the density matrix is separable. While the use of Eq. (1.14) as a criterion for separability is quite simple one should take into account that it is restricted to the special cases in which $A$ or $B$ are equal to zero.

### 3. Lorentz transformations for symmetric two qubits

We simplify the analysis in this section by studying the symmetric case i.e. $a_i = b_i$ (i=1, 2, 3). This condition is valid for cases in which the analysis does not distinguish between the two qubits and simplifies very much the analysis. For the symmetric case, $R$ can be written in the form

$$R = \begin{pmatrix} 1 & a_1 & a_2 & a_3 \\ a_1 & t_1 & 0 & 0 \\ a_2 & 0 & t_2 & 0 \\ a_2 & 0 & 0 & t_3 \end{pmatrix} \quad . \quad . \tag{3.1}$$

The density matrix corresponding to $R$ of Eq. (3.1) is given by

$$4\rho = \begin{pmatrix} 1 + 2a_3 + t_3 & a_1 - a_2 i & a_1 - a_2 i & t_1 - t_2 \\ a_1 + a_2 i & 1 - t_3 & t_1 + t_2 & a_1 - a_2 i \\ a_1 + a_2 i & t_1 + t_2 & 1 - t_3 & a_1 - a_2 i \\ t_1 - t_2 & a_1 + a_2 i & a_1 + a_2 i & 1 - 2a_3 + t_3 \end{pmatrix} \quad . \tag{3.2}$$

It is interesting to note that one of the eigenvalues is given by $4\lambda = 1 - t_1 - t_2 - t_3$.

The PT transformation is given by

$$4\rho(PT) = \begin{pmatrix} 1+2a_3+t_3 & a_1-a_2 i & a_1+a_2 i & t_1+t_2 \\ a_1+a_2 i & 1-t_3 & t_1-t_2 & a_1+a_2 i \\ a_1-a_2 i & t_1-t_2 & 1-t_3 & a_1-a_2 i \\ t_1+t_2 & a_1-a_2 i & a_1+a_2 i & 1-2a_3+t_3 \end{pmatrix}. \qquad (3.3)$$

For cases for which $\rho$ represents a density matrix (i.e. all its eigenvalues are non-negative), $\rho$ will be separable if and only if all the eigenvalues of $\rho(PT)$ are non-negative. This "existence" theorem does not tell us, however, the explicit separable form of the density matrix.[10] For this purpose we would like to eliminate the parameters $a_i (i=1,2,3)$ by using the Lorentz transformation which transforms $R$ into $Q$ given by[5]

$$Q = L \cdot R \cdot L^\dagger . \qquad (3.4)$$

The general matrix $L$ is[18]

$$L = \begin{pmatrix} \gamma & -\gamma\beta_1 & -\gamma\beta_2 & -\gamma\beta_3 \\ -\gamma\beta_1 & 1+\beta_1^2 X & \beta_1\beta_2 X & \beta_1\beta_3 X \\ -\gamma\beta_2 & \beta_1\beta_2 X & 1+\beta_2^2 X & \beta_2\beta_3 X \\ -\gamma\beta_3 & \beta_1\beta_3 X & \beta_2\beta_3 X & 1+\beta_3^2 X \end{pmatrix} \qquad (3.5)$$

Here

$$X = \frac{\gamma-1}{\beta^2} \; ; \; \beta^2 = \beta_1^2 + \beta_2^2 + \beta_3^2 \; ; \; \gamma = (1-\beta^2)^{-1/2} . \qquad (3.6)$$

In the present analysis $L = L^\dagger$ since all parameters $a_1, a_2, a_3, t_1, t_2, t_3$, are real and $R$ is assumed to be symmetric. The Lorentz transformation eliminates the parameters $a_i$ $(i=1,2,3)$ by the requirements

$$Q_{0,1} = Q_{1,0} = -\gamma\beta_1(\gamma - a_1\gamma\beta_1 - a_2\gamma\beta_2 - a_3\gamma\beta_3) + \beta_1\beta_2(a_2\gamma - \gamma\beta_2 t_2)X + \\ \beta_1\beta_3(a_3\gamma - \gamma\beta_3 t_3)X + (a_1\gamma - \gamma\beta_1 t_1)(1+\beta_1^2 X) = 0 \qquad (3.7)$$

$$Q_{0,2} = Q_{2,0} = -\gamma\beta_2(\gamma - a_1\gamma\beta_1 - a_2\gamma\beta_2 - a_3\gamma\beta_3) + \beta_1\beta_2(a_1\gamma - \gamma\beta_1 t_1)X + \\ \beta_2\beta_3(a_3\gamma - \gamma\beta_3 t_3)X + (a_2\gamma - \gamma\beta_2 t_2)(1+\beta_2^2 X) = 0 \qquad (3.8)$$

$$Q_{0,3} = Q_{3,0} = -\gamma\beta_3(\gamma - a_1\gamma\beta_1 - a_2\gamma\beta_2 - a_3\gamma\beta_3) + \beta_1\beta_3(a_1\gamma - \gamma\beta_1 t_1)X + \\ \beta_2\beta_3(a_2\gamma - \gamma\beta_2 t_2)X + (a_3\gamma - \gamma\beta_3 t_3)(1+\beta_3^2 X) = 0 \qquad (3.9)$$

The left hand side of Eq. (3.7) multiplied by $\beta_2$ minus the left hand side of Eq. (3.8) multiplied by $\beta_1$ gives after a straightforward calculation the relation

$$\frac{a_1 - \beta_1 t_1}{\beta_1} = \frac{a_2 - \beta_2 t_2}{\beta_2} \quad . \tag{3.10}$$

The left hand side of Eq. (3.7) multiplied by $\beta_3$ minus the left hand side of Eq. (3.8) multiplied by $\beta_2$ gives after a straightforward calculation the relation

$$\frac{a_2 - \beta_2 t_2}{\beta_2} = \frac{a_3 - \beta_3 t_3}{\beta_3} \quad . \tag{3.11}$$

Equations (3.10-3.11) give the relations

$$\beta_1 \beta_2 = \frac{a_2 \beta_1 - a_1 \beta_2}{t_2 - t_1} \quad ; \quad \beta_2 \beta_3 = \frac{a_3 \beta_2 - a_2 \beta_3}{t_3 - t_2} \quad ; \quad \beta_3 \beta_1 = \frac{a_3 \beta_1 - a_1 \beta_3}{t_1 - t_3} \quad , \tag{3.12}$$

and the simple relations:

$$\beta_2 = \frac{a_2 \beta_1}{a_1 + \beta_1 (t_2 - t_1)} \quad ; \quad \beta_3 = \frac{a_3 \beta_1}{a_1 + \beta_1 (t_3 - t_1)} \quad . \tag{3.13}$$

Multiplying Eq. (3.7) by $\beta_2 \beta_3$, Eq. (3.8) by $\beta_1 \beta_3$, and Eq. (3.9) by $\beta_2 \beta_3$, adding these 3 equations and substituting in these equations the relations (3.10, 3.11) we get after straightforward calculations the relation

$$\frac{a_1 - \beta_1 t_1}{\beta_1} = 1 - a_1 \beta_1 - a_2 \beta_2 - a_3 \beta_3 \quad . \tag{3.14}$$

Substituting the relations (3.13) into Eq. (3.14) leads to the equation

$$a_1 - \beta_1 t_1 = \beta_1 - a_1 \beta_1^2 - a_2^2 \left( \frac{\beta_1^2}{a_1 + \beta_1 (t_2 - t_1)} \right) - a_3^2 \left( \frac{\beta_1^2}{a_1 + \beta_1 (t_3 - t_1)} \right) \quad . \tag{3.15}$$

A physical solution of the above equations (i.e. $\beta^2 < 1$) results in $Q_{0,i} = Q_{i,0} = 0 \; (i=1,2,3)$, and $Q_{0,0} = s_0$. The resulting $Q_{i,j} \; (i,j = 1,2,3)$ is symmetric and can be diagonalized by 3-dimensional rotation. If there is no physical solution for $\beta$ it means that $\rho$ belongs to the non-generic cases.

The analysis is simplified under the condition that one of the parameters $a_1, a_2, a_3$ vanishes, e.g. $a_3 = 0$. Then by introducing the definitions

$$t = t_2 - t_1 \quad ; \quad T = 1 + t_1 \quad , \tag{3.16}$$

Eq. (3.15) is reduced to the cubic equation:

$$\frac{a_1}{t} + \beta_1\left(1 - \frac{T}{t}\right) + \frac{\beta_1^2}{a_1}\left(\frac{a_1^2 + a_2^2}{t} - T\right) + \beta_1^3 = 0 \quad . \tag{3.17}$$

Eq. (3.15), including non-vanishing 3 symmetric parameters $a_1, a_2, a_3$, and using the definitions (3.16) and the additional definition $t_3 - t_1 = t'$, can be transformed to a quartic equation given as:

$$\beta_1^4 + \beta_1^3\left[\frac{a_1}{t} + \frac{a_1}{t'} - \frac{T}{a_1} + \frac{a_2^2}{a_1 t} + \frac{a_3^2}{a_1 t'}\right] +$$
$$\beta_1^2\left[1 + \frac{a_1^2 + a_2^2 + a_3^2}{tt'} - \frac{T}{t'} - \frac{T}{t}\right] + \beta_1\left[\frac{a_1}{t'} + \frac{a_1}{t} - a_1\frac{T}{tt'}\right] + \frac{a_1^2}{tt'} = 0 \tag{3.18}$$

The solution of the cubic equation (3.17) and the quartic equation (3.18) can be made by conventional methods.[19] The analytical results become quite complicated as they include complicated functions of the parameters $T, t, a_1, a_2, a_3$ and $t'$ but for any numerical values of these parameters the cubic and quartic equations can be easily solved. One should take care that such real solution satisfies the relation $\beta_1^2 + \beta_2^2 + \beta_3^2 < 1$.

We demonstrate our analysis by using the Lorentz transformation in an example where the parameters of $R$ Eq. (3.1)) are: $a_1 = 0.1; a_2 = 0.15, t_1 = 0.3; t_2 = -0.2; t_3 = 0.4; a_3 = 0$. The matrix $R$ in this example is given by

$$R = \begin{pmatrix} 1 & 0.1 & 0.15 & 0 \\ 0.1 & 0.3 & 0 & 0 \\ 0.15 & 0 & -0.2 & 0 \\ 0 & 0 & 0 & 0.4 \end{pmatrix} \quad . \tag{3.19}$$

As the eigenvalues of $\rho(PT)$ are non-negative we conclude that this density matrix is separable but in order to construct an explicitly separable form for the density matrix[10] we eliminate the linear terms by a Lorentz transformation. Inserting the above parameters in Eq. (3.17) we get the cubic equation:

$$\beta_1^3 - 13.65\beta_1^2 + 3.6\beta_1 - 0.2 = 0 \quad . \tag{3.20}$$

While this equation gives 3 solutions for $\beta_1$ only the smallest one given by $\beta_1 = 0.0792$ is valid physically. (We need to choose the solution with the smallest value of $\beta$ so that $\beta$ will vanish in the limit $a_i \to 0$ $(i = 1, 2, 3)$. A similar explanation is given in the comment after Eq. (2.10).)

Using Eq. (3.13) we get in this example: $\beta_2 = 0.1967$ (and $\beta^2 = \beta_1^2 + \beta_2^2 < 1$).

We substitute $\beta_1 = 0.0792$, $\beta_2 = 0.1967$ and $\beta_3 = 0$ in Eq. (3.5), and using the relations (3.6) we obtain the Lorentz matrix $L$. Then by substituting this matrix and Eq. (3.19) into Eq. (3.4) the numerical calculations in this example give:

$$Q = L \cdot R \cdot L^{\dagger} = \begin{pmatrix} 0.96257 & 0 & 0 & 0 \\ 0 & 0.292049 & -0.015808 & 0 \\ 0 & -0.015808 & -0.229474 & 0 \\ 0 & 0 & 0 & 0.4 \end{pmatrix}. \quad (3.21\}$$

One should notice that the linear terms were eliminated by the Lorentz transformation. As the matrix $Q$ is symmetric we calculate the eigenvalues of (3.21) and get

$$s_1/s_0 = 0.303945 \quad ; \quad s_2/s_0 = -0.238396 \quad ; \quad s_3/s_0 = 0.415552 \quad . \quad (3.22)$$

Since $|s_1/s_0| + |s_2/s_0| + |s_3/s_0| = 0.957893 < 1$ this density matrix is separable, in consistence with the positivity of the eigenvalues of $\rho(PT)$.

We demonstrate the use of Lorentz transformations with 3 $\beta's$ in another example in which R of (3.1) has the parameters: $a_1 = 0.1, a_2 = 0.15, a_3 = 0.2, t_1 = 0.3, t_2 = -0.2, t_3 = 0.2$. The matrix $R$ in this example is given by

$$R = \begin{pmatrix} 1 & 0.1 & 0.15 & 0.2 \\ 0.1 & 0.3 & 0 & 0 \\ 0.15 & 0 & -0.2 & 0 \\ 0.2 & 0 & 0 & 0.2 \end{pmatrix} . \quad (3.23)$$

The eigenvalues of $\rho$ are $\lambda_1 = 1.8894; \lambda_2 = 1.0256; \lambda_3 = 0.7; \lambda_4 = 0.3850$, and those of $\rho(PT)$ are $\lambda_1 = 1.7181; \lambda_2 = 1.2638; \lambda_3 = 0.8195; \lambda_4 = 0.1986$. As all the eigenvalues of $\rho(PT)$ are non-negative we conclude that this density matrix is separable but in order to construct an explicitly separable form for the density matrix [10] we would like to eliminate the linear terms by the use of Lorentz transformations. Inserting the above parameters in Eq. (3.18) we get the quartic equation:

$$\beta_1^4 - 18.63\beta_1^3 + 18.05\beta_1^2 - 3.8\beta_1 + 0.2 = 0 \quad . \quad (3.24)$$

While this equation gives 4 solutions for $\beta_1$ we use only the smallest one given by $\beta_1 = 0.0816..$
By using Eq. (3.13) we get in this example $\beta_2 = 0.2068$ ; $\beta_3 = 0.1777$ ( $\beta_1^2 + \beta_2^2 + \beta_3^2 < 1$ ).

Substituting $\beta_1 = 0.0816$, $\beta_2 = 0.2068$, $\beta_3 = 0.1777$, in Eq. (3.5), and using the relations (3.6) we obtain the Lorentz matrix $L$. Then by substituting this matrix and Eq. (3.23) in Eq. (3.4), the calculations in this example give:

$$Q = L \cdot R \cdot L^\dagger = \begin{pmatrix} 0.92527 & 0 & 0 & 0 \\ 0 & 0.29218 & -0.01648 & -0.01713 \\ 0 & -0.01648 & -0.230845 & -0.03401 \\ 0 & -0.01713 & -0.03401 & 0.16432 \end{pmatrix} . \quad (3.25)$$

As the matrix $Q$ is symmetric we calculate the eigenvalues of (3.25) and get

$$s_0 = 0.9257 \quad ; \quad s_1 = 0.2943 \quad ; \quad s_3 - 0.2344 \quad ; \quad s_4 = 0.1653 . \quad (3.26)$$

The condition (1.16) for separability is satisfied, and we have the explicit separable form.[10]

## 4. Discussion, summary and conclusions

In the present work we studied certain properties of two qubits density matrices which include linear terms and can be given in the form of Eq. (1.3). This density matrix can be transformed into the form of Eq. (1.4) in which the matrix $R_{\mu,\nu}$ is of $4\times 4$ dimension. The properties of $R_{\mu,\nu}$ and the corresponding two qubits density matrices under local transformations have been analyzed in many works.[3-12] The Peres-Horodecki criterion gives a necessary and sufficient condition for separability of two qubits density matrices but does not give the explicitly separable form as function of the parameters $t_i, a_i, b_i$ ($i = 1, 2, 3$). It has been shown in various works[10,13-15] that under the condition $\vec{a} = \vec{b} = 0$ a necessary and sufficient condition for separability is given by $\sum_{i=1}^{3} |t_i| \leq 1$, and an explicitly separable form was given.[10] For a two qubits density matrix which includes linear terms this condition is necessary but not sufficient. Verstraete et al.,[4-6] have shown that it is possible to diagonalize $R_{\mu,\nu}$ by Lorentz transformations, in most cases referred as the generic cases, so that sufficient and necessary condition for separability is given by Eq. (1.16), or alternatively in normalized form of Eq. (1.18). Since Lorentz transformations are not orthogonal there are also cases in which $R_{\mu,\nu}$ cannot be diagonalized, referred to as the non-generic cases. While Verstraete et al. [4-6] gave important existence theorems about the application of Lorentz transformations, they did not give explicit relations between the parameters $t_i, a_i, b_i$ $(i = 1, 2, 3)$ and $s_i$ $(i = 0, 1, 2, 3)$. In the present work we developed such relations for certain important generic cases, and showed explicitly why the non-generic cases cannot be cast in diagonal form by Lorentz transformations.

In section 2 we used Lorentz transformations for the density matrix of two qubits which includes only one pair of linear terms. The density matrix for this case is given by Eq. (2.2) and the eigenvalues by Eq. (2.3). By using Lorentz transformations, $R$ of Eq. (2.1) was transformed to $R_L$ of Eq. (2.5) which includes $\beta_A$ and $\beta_B$. For the generic case the non-diagonal elements of $R_L$ were eliminated. We treated two limiting cases of Eq. (2.10). For the case in which $a_1 = b_1 = a$, in Eq. (2.11), we have $\beta_A = \beta_B = \beta$. Then the necessary and sufficient condition for separability is given by Eq. (2.14). We demonstrated the use of this equation in an example. For the extreme non-symmetric case in which $b_1 = 0$, $\beta_A$ is given by Eq. (2.15), and the necessary and sufficient condition for separability is given by Eq. (2.21). We demonstrated the use of this equation in an example. For the four non–generic cases [6] treated by us as cases $a), b), c), d)$, we showed that one of the $\beta's$ equals 1 so that diagonalization is not reachable physically.

In section 3 we applied Lorentz transformations for symmetric two qubits including three (or two) pairs of linear terms, obtaining a quartic (or cubic) equation. The matrix $R$ is given by Eq. (3.1). The density matrix $\rho$ and its PT transformation were given, respectively, by Eq. (3.2) and (3.3) and the Lorentz matrix $L$ is given by Eq. (3.5). The Lorentz transformation eliminates the linear terms $a_1, a_2, a_3$, by using, respectively, Eq.(3.7), (3.8) and (3.9). From these equations $\beta_2$ and $\beta_3$ were given as functions of $\beta_1$ in Eq. (3.13) and the fundamental equation for $\beta_1$ was given in Eq. (3.14). For the case in which we have the parameters $a_1, a_2, t = t_2 - t_1, T = 1 + t_1, t_3, a_3 = 0$, we get the cubic equation (3.17). For the case in which we have $a_1, a_2, a_3, t = t_2 - t_1, T = 1 + t_1, t' = t_3 - t_1$ we get the quartic equation (3.18). We demonstrated the use of the cubic equation in an example in which $\beta_1$ and $\beta_2$ were calculated. Then by using the Lorentz transformation the linear terms were eliminated in Eq. (3.21). By calculating the eigenvalues in this symmetric case we derived $s_\mu$ $(\mu = 0, 1, 2, 3)$ in Eq. (3.22). We demonstrated the use of the quartic equation in an example in which $\beta_1, \beta_2$ and $\beta_3$ were calculated. Then by using Lorentz transformation the linear terms were eliminated in Eq. (3.25). By calculating the eigenvalues in this symmetric case we derived $s_\mu$ ($\mu = 0,1,2,3$) in Eq. (3.26). The crucial point in our derivations is that we find explicit expressions for these parameters.